\documentclass[twocolumn,showpacs,pra,aps]{revtex4}
\usepackage{array}
\usepackage{amsmath}
\usepackage{graphicx}

\begin{document}

\title{Entanglement of a two-mode squeezed state in a phase-sensitive Gaussian
environment}

\author{D. Wilson}

\author{Jinhyoung Lee}

\author{M. S. Kim}

\affiliation{School of Mathematics and Physics, Queen's University,
  Belfast BT7 1NN, United Kingdom}

\date{\today}

\begin{abstract}
 Some non-classical properties such as squeezing, sub-Poissonian photon statistics
or oscillations in photon-number distributions may survive longer in a
phase-sensitive environment
than in a phase-insensitive environment.  We examine if entanglement, which is an inter-mode
non-classical feature, can also survive longer in a phase-sensitive environment.  Differently
from the single-mode case, we find that making the environment phase-sensitive does not
aid in prolonging the inter-mode non-classical nature, {\em i.e.}, entanglement.

\end{abstract}

\maketitle

\newpage
\section{INTRODUCTION}

\noindent

There have been an increasing number of discussions on entanglement and its applications
in the context of quantum information processing.  In particular, because of convenient
manipulation of lasers, entanglement for Gaussian continuous-variable states has been
studied extensively \cite{Braunstein-book}.

A two-mode squeezed state is an entangled Gaussian continuous-variable
state which becomes
maximally entangled when it is infinitely squeezed.  Once the pure two-mode squeezed
state is embedded in an environment, entanglement is degraded.  In this paper, we are interested in
the dynamics of  entanglement in an environment.
In the vacuum environment,
the two-mode squeezed state never becomes separable regardless of the initial degree
of squeezing \cite{duan-cirac}.  At non-zero temperature, entanglement is degraded faster and
it is eventually lost \cite{duan-cirac,kim,scheel}.

In quantum optics, correlated (phase-sensitive) reservoirs \cite{dupertuisbarnettstenholm},
sometimes called ``rigged reservoirs'', based on the establishment of squeezed
light have been studied extensively.  It has been shown that the decay
rate of quantum coherence
for a single-mode field can be significantly modified compared to the decay rate in an ordinary
phase-insensitive thermal reservoir \cite{Kennedy,kim-buzek}.
Depending on the phases of the quantum system and the correlated
reservoir the decay rate of quantum coherence can be either enhanced or significantly
suppressed.  Suppression of the decay rate of the quantum coherence leads to
preservation of non-classical features such as squeezing, sub-Poissonicity and
oscillations in photon number distribution
\cite{kim-buzek}.

In this paper, we answer the question ``Will the phase sensitivity of the environment
preserve the entanglement nature of a two-mode squeezed state?''.  Entanglement is
due to inter-mode quantum coherence.  Is there a possibility to keep the inter-mode
quantum coherence as we squeeze the environment?  To answer this question, we first find
the dynamics of the two-mode squeezed state in a phase-sensitive environment in Section II.
We then analyse the entanglement of this state using the entanglement condition
developed for Gaussian continuous-variable states \cite{simon} in
Section III.  The dynamics of entanglement are considered when one
mode of the environment is phase-sensitive while the other is 
phase-insensitive in Section IV.

\section{Dynamics of Two-mode Squeezed State in a Gaussian Environment}
There is a one-to-one correspondence between the density operator of a field
and its Weyl characteristic function.
The use of characteristic functions and quasi-distribution functions
\cite{barnettknight}
allow the calculation of the expectation values of functions of the
annihilation and creation operators.  Characteristic functions and
quasi-distribution functions are complete descriptions of a field as
they contain all the information necessary to reconstruct the density
matrix.  There are a group of quasi-probability functions among which
P, Q and Wigner functions are more frequently used \cite{Glauber}.

The Weyl characteristic function, which is the inverse Fourier transform of the
Wigner function, is defined for a two-mode field with its density operator
$\hat{\rho}$ as
\begin{equation}
\label{character}
C^{(W)}(\eta,\xi) = {\rm Tr}[\hat{\rho}\hat{D}(\eta)\hat{D}(\xi)]
\end{equation}
where $\hat{D}(\eta)={\rm exp}(\eta\hat{a}^\dagger - \eta^*\hat{a})$
is the Glauber displacement operator \cite{barnettknight}.

It has been found by one of us \cite{kim-imoto} that the
interaction of a field with a phase-sensitive environment can be
modelled by the field passing through an array composed of an infinite
number of beam splitters. The physical properties of the environment are
determined by the fields injected into unused ports of the beam
splitters. Using this model, it has been found that the dynamic
characteristic function $C^{(W)}(\eta,\xi,\tau)$ for the system
interaction with the environment, is represented by
\begin{equation}
\label{dynamic-chara} C^{(W)}(\eta,\xi,\tau) = C_e^{(W)}(r\eta,
r\xi)C_s^{(W)}(t\eta, t\xi)
\end{equation}
where $e$ denotes the environment and $s$ the system, $r$ is the
normalised dimensionless time $r = \sqrt{1 -
\rm{exp}(-\gamma\tau)}$, $\gamma$ is the coupling between the
environment and system, $\tau$ is the interaction time and $t =
\sqrt{1-r^2}$.  Kim and Imoto \cite{kim-imoto} rigorously proved
for a single-mode field that this is a solution of the
Fokker-Planck equation normally encountered in the problem of a
system interacting with a Gaussian environment. If the two-modes
of the environment are independent of each other, which is what
we are interested in in this paper, the characteristic function
for the environment is the product of the characteristic functions
for each environment:
\begin{equation}
C_e^{(W)}(r\eta, r\xi)=C_{e1}^{(W)}(r\eta)C_{e2}^{(W)}(r\xi).
\end{equation}

A two-mode squeezed state is defined as $\hat{S}_{ab}(s_c) \left |
0_A 0_B \right >$ where the two-mode squeezing operator
$\hat{S}_{ab}(s_c) = \exp(s_c\hat{b}\hat{a} - s_c
\hat{a}^\dagger\hat{b}^\dagger)$ and $\hat{a}$ and $\hat{b}$ are
the field bosonic operators \cite{barnettknight}.  For simplicity,
we have assumed the squeezing parameter $s_c$ real and positive.
Using its definition (\ref{character}), the Weyl characteristic
function is found for the two-mode squeezed state:
\begin{equation}
C^{(W)}_s(\eta, \xi) = {\rm exp}\left[-\frac{1}{2}(\eta ~ \xi){\bf
V}(\eta ~ \xi)^T\right]
\end{equation}
where ${\bf V}$ is the variance matrix
\begin{equation}
{\bf V} = \left (
\begin{array}{cc}
\mu \openone & -\lambda {\boldsymbol \sigma}_z  \\
-\lambda \boldsymbol{\sigma}_z &  \mu \openone \\
\end{array}
\right )
\end{equation}
with $\mu = \cosh 2s_c$, $\lambda = \sinh 2s_c$ and the
Pauli spin matrix ${\boldsymbol \sigma}_z$. Throughout the paper, a matrix is
represented in bold face and an operator by a hat.

In order to find the dynamic characteristic function
$C(\eta,\xi,\tau)$ for the interaction with the phase-sensitive
environment, we use Eq.(\ref{dynamic-chara}) with $C_e(\eta,\xi)$
for the squeezed thermal states of modes $a$ and $b$
\cite{kim-imoto}. Here, the Weyl characteristic function for the
squeezed thermal state of mode $a$ is  $C^{(W)}_{e1}(\eta) = {\rm
Tr}[\hat{S}(\zeta_{e1})\hat{\rho}_{th}\hat{S}^\dagger(\zeta_{e1})\hat{D}(\eta)]$
where the single-mode squeezing operator
$\hat{S}(\zeta)=\exp[\frac{1}{2}(\zeta^*\hat{a}^2 - \zeta
(\hat{a}^\dagger)^2]$ and $\hat{\rho}_{th}$ is the density
operator for the thermal state of average photon number $\bar n$.
We keep the squeezing parameter $\zeta_{e1}=s_{e1}\exp(\phi_{e1})$
complex to find the optimum condition for the environment to maintain
the entanglement of the system. After a little calculation, the
Weyl characteristic function is found:
\begin{equation}
\label{single-char} C^{(W)}_{e1}(\eta)= {\rm
exp}\left[-\frac{1}{2}(\eta_i ~ \eta_r) {\bf R}(\eta_i ~
\eta_r)^T\right]
\end{equation}
where $\eta_i$ and $\eta_r$ are imaginary and real parts of $\eta$
and the single-mode variance matrix
\begin{displaymath}
{\bf R} =
\left (
\begin{array}{cc}
a_- & b   \\
b   & a_+ \\
\end{array}
\right )
\end{displaymath}
with $a_\pm = \tilde{n}(\cosh 2s_{e1} \pm \cos \phi_{e1} \sinh
2s_{e1})$ and $b=\tilde{n}\sin\phi_{e1}\sinh 2s_{e1}$ with
$\tilde{n}=2\bar{n}+1$.  The average excitation of the environment
is
\begin{equation}
\label{average} <\hat{a}^\dagger\hat{a}> = \bar{n}\cosh 2s_{e1}
+ \sinh^2 s_{e1}.
\end{equation}
When $\zeta_{e1}=0$ the phase-sensitive environment becomes a
phase-insensitive thermal bath of average photon number
$\bar{n}$.  Similarly, we can obtain $C^{(W)}_{e2}(\xi)$ for the
environment of mode $b$.  Throughout the paper $s_{e1}$ and $s_{e2}$
are taken positive without losing generality as $\phi_{e1}$ and $\phi_{e2}$
are not fixed.

Substituting $C^{(W)}_{e1}(\eta)$ and $C^{(W)}_{e2}(\xi)$ into
Eq.(\ref{dynamic-chara}), the dynamic characteristic function is
found:
\begin{equation}
\label{Wigner-Weyl}
C^{(W)}(\eta, \xi, \tau) = {\rm exp} \left
[-\frac{1}{2}(\eta_i~\eta_r~\xi_i~\xi_r){\bf V'}
(\eta_i~\eta_r~\xi_i~\xi_r)^T\right ]
\end{equation}
where ${\bf V'} = t^2{\bf V} + r^2{\bf R}\otimes\openone$.

\section{Separability of the dynamic field}
Simon \cite{simon} found the sufficient and necessary condition
for the separability of a Gaussian continuous-variable state. Simon
found the condition based on the fact that the partial transpose
of its density operator should be positive if a two-mode state is
separable \cite{peres}. A two-mode Gaussian state is represented
by its variance matrix
\begin{equation}
\label{gaussian} {\bf V}_G = \left (
\begin{array}{cc}
{\bf A} & {\bf C} \\
{\bf C}^T & {\bf B} \\
\end{array}
\right )
\end{equation}
where ${\bf A}$, ${\bf B}$ and ${\bf C}$ are $2\times2$ matrices
and Simon's separability condition becomes
\begin{eqnarray}
\label{SimonsCriteriondetform} 
& ({\rm det}{\bf A} - 1)({\rm
det}{\bf B} - 1)+(|{\rm det}{\bf C}| - 1)^2-1 & \nonumber \\ 
& \geq{\rm Tr}[{\bf
A}{\boldsymbol \sigma}_y {\bf C} {\boldsymbol \sigma}_y {\bf
B}{\boldsymbol \sigma}_y {\bf C}^T{\boldsymbol \sigma}_y] &
\end{eqnarray}
where ${\boldsymbol \sigma}_y$ is the Pauli matrix. When the criterion is
satisfied the system is separable.

Assuming the two independent modes of the reservoir are prepared
in the same condition, {\it i.e.} $\zeta_{e1} = \zeta_{e2} \equiv
\zeta_e$, the variance matrix of the channel interacting with this
environment is
\begin{displaymath}
{\bf V'} = \left (
\begin{array}{cc}
\mu t^2\openone+r^2{\bf R} & -\lambda t^2{\boldsymbol \sigma}_z  \\
-\lambda t^2{\boldsymbol \sigma}_z & \mu t^2\openone+r^2{\bf R}  \\
\end{array}
\right )
\end{displaymath}
For this variance matrix, the left-hand side (LHS) of Simon's
criterion (\ref{SimonsCriteriondetform}) is
\begin{equation}
\left[t^4\mu^2 + r^4\tilde{n}^2 + 2r^2t^2\mu\tilde{n}\cosh
2s_e-1\right]^2+ (\lambda^2t^4-1)^2-1,
\end{equation}
which clearly shows that the LHS of the separability criterion is
independent of the phase $\phi$ of the environment squeezing. On
the other hand, the right-hand side (RHS) of the criterion is
phase-dependent:
\begin{eqnarray}
& 2\lambda\mu t^4(\mu t^2 + 2\tilde{n}r^2\cosh 2s_e) & \nonumber \\
& + 2\lambda t^2r^4\tilde{n}^2(\cosh^2 2s_e + \cos 2\phi_e \sinh^2 2s_e), &
\end{eqnarray}
which is maximised when $\phi_e = 0$.  In other words, the
lifetime of entanglement is maximised for $\phi_e=0$, which is
the same squeezing angle as that for the two-mode squeezed state
of the system (Recall $s_c$ has been taken positive real).

Now we want to examine the possibility that the entanglement of
the two-mode squeezed state can survive longer by squeezing the
environment.  Before we do this, we find that the separability
condition becomes extremely simple when $\phi_{e}=0$ as the
following Lemma shows.

{\em Lemma 1.} --- If the block matrices ${\bf A}={\bf B}$ and
${\bf
  C}$ are diagonal, {\em i.e.}, the quadrature matrix of a Gaussian
continuous-variable state has the following form:
\begin{equation}
{\bf V}_o= \left (
\begin{array}{cccc}
  n_1 & 0 & c_1 & 0 \\
  0 & n_2 & 0 & c_2 \\
  c_1 & 0 & n_1 & 0 \\
  0 & c_2 & 0 & n_2
\end{array}
\right )
\label{standard-0}
\end{equation}
where $n_1$ or $n_2$ may be smaller than the vacuum limit 1. The
state is separable if and only if
\begin{equation}
\label{separability-condition} (n_1-|c_1|)(n_2-|c_2|)\geq 1.
\end{equation}

{\em Proof} --- By local unitary squeezing operations, the matrix
(\ref{standard-0}) is transformed into
\begin{equation}
{\bf V}_1=\left (
\begin{array}{cccc}
  n & 0 & c & 0 \\
  0 & n & 0 & c^\prime \\
  c & 0 & n & 0 \\
  0 & c^\prime & 0 & n \\
\end{array}\right )
\label{standard-1}
\end{equation}
where $n=\sqrt{n_1 n_2}$, $c=c_1\sqrt{n_2/n_1}$ and
$c^\prime=c_2\sqrt{n_1/n_2}$.  The factor $n$ is directly related
to the uncertainty principle to satisfy $n\geq 1$.  For the state
with the quadrature matrix (\ref{standard-1}), Simon's
separability condition \cite{simon} reads
\begin{equation}
\label{Simon-condition} (n^2-c^2)(n^2-c^{\prime 2})\geq
2n^2+2|cc^\prime|-1.
\end{equation}
Define the mean and the difference of the correlation factors,
$|c|$ and $|c^\prime|$: $c_m=(|c|+|c^\prime|)/2$ and
$c_d=(|c|-|c^\prime|)/2$. Using the new parameters $c_m$ and
$c_d$, Simon's criterion (\ref{Simon-condition}) can be written as
\begin{equation}
\label{Simon-factor}
[(n-c_m)^2-(1+c_d^2)][(n+c_m)^2-(1+c_d^2)]\geq 0
\end{equation}
where $[(n+c_m)^2-(1+c_d^2)]$ is positive unless $n=1$ and
$c=c^\prime=0$ when it becomes zero.  The separability condition
is satisfied if and only if
\begin{equation}
\label{Simon-2}
 (n-c_m)^2-(1+c_d^2)\geq 0 \Leftrightarrow (n-|c|)(n-|c^\prime|)\geq 1.
\end{equation}
With use of the definitions of $n$, $c$ and $c^\prime$ for the
inequality on the RHS of the arrow we obtain the separability
condition in Eq.~(\ref{separability-condition}).

When $\phi_e=0$ we can use the separability condition
(\ref{separability-condition}) which leads to
\begin{equation}
\label{separable}
(t^2\mbox{e}^{-2s_c}+r^2\tilde{n}\mbox{e}^{2s_e})(t^2\mbox{e}^{-2s_c}+r^2\tilde{n}\mbox{e}^{-2s_e})\geq
1.
\end{equation}
It is obvious that the LHS is minimised when $s_e=0$ for fixed $t,
s_c$ and $\tilde{n}$. We have proved that {\em squeezing the
environment does not help to keep the entanglement of the two-mode
squeezed state longer}.  This is one of the main results of this paper.

Entanglement, which is due to the inter-mode coherence, does not
survive longer in the phase-sensitive than in the phase-insensitive
environment.  This result contrasts to the single-mode case where
quantum coherence may last longer in the phase-sensitive
environment.  A system always gains extra noise due to its
interaction with an environment.  For a phase-sensitive
environment, the extra noise is reduced along one
quadrature axis at the expense of expanding noise along its
conjugate axis.  The quantum coherence imposed in one quadrature
can last longer by adding the reduced noise.  In the variance
matrix (\ref{standard-0}), $n_1$ and $n_2$ determine the
quadrature variances for local states of modes $a$ and $b$, and $c$
and $c^\prime$ show the inter-mode correlation
\cite{simon,kim-lee-munro}.  As the initial two-mode squeezed
state $n_1=n_2$, interacts with an environment, $n_1$ and
$n_2$ grow.  If we squeeze the environment, the amounts of added
noise become different and the LHS of Eq.
(\ref{separability-condition}) gets larger.  We have seen that
because the entanglement is due to both quadrature variables,
reducing the extra noise in one quadrature does not lengthen its
lifetime.  In fact, we can interpret this using the Einstein-Podolsky-Rosen (EPR)
argument \cite{EPR}, which is related to entanglement.  EPR argues that
it is possible to infer values for two conjugate variables
of particle 2 from the measurement of their counter parts for particle 1
when the two particles are perfectly correlated.  The EPR argument is clearly concerned with the product of uncertainties for conjugate quadratures which is increased
more rapidly by interaction with phase-sensitive reservoirs.

\section{Different conditions for two environment modes}
We have so far considered the case where the two modes of the
environment are in the same condition.  By this restriction, the problem became simple and
we could see
the impact of squeezing the environment clearly.  What happens if the two
modes are under different conditions?  This case was considered
for phase-insensitive environments \cite{kim,scheel}.  When the
two modes of the environments are in different conditions, we
are not able to use the separability condition
(\ref{separability-condition}) because the variance matrix is no
longer in the form (\ref{standard-0}).  

The consideration of very general interaction with phase-sensitive
environments is complicated so we still restrict ourselves to the case
where $\phi_{e1}=\phi_{e2}=0$, $\bar{n}_{e1}=\bar{n}_{e2}$ and
normalised interaction times $r_1=r_2$.
The dynamic characteristic function for the interaction of the system
with the two environments with different squeezing is represented by the
variance matrix
\begin{displaymath}
{\bf V'} = \frac{1}{2} \left (
\begin{array}{cccc}
n_1 & 0 & c_1 & 0 \\
0 & n_2 & 0 & c_2 \\
c_1  & 0 & m_1 & 0 \\
0 & c_2 & 0 & m_2
\end{array}
\right ),
\end{displaymath}
where
\begin{eqnarray}
n_1 =  t^2\mu+r^2\tilde{n}\mbox{e}^{-2s_{e1}} &~;~& n_2  = t^2\mu
+ r^2\tilde{n}\mbox{e}^{2s_{e1}}\nonumber \\
m_1  =  t^2\mu + r^2\tilde{n}\mbox{e}^{-2s_{e2}} &~;~& m_2  = t^2\mu
+ r^2\tilde{n}e^{2s_{e2}}
\label{19}
\end{eqnarray}
and $|c_1|=|c_2|=\lambda t^2$.  Simon's separability criterion (\ref{SimonsCriteriondetform}) then
becomes
\begin{eqnarray}
\label{OurSimonsCriterion} 
& \delta \equiv (t^4\lambda^2-1)^2 -t^4\lambda^2(n_1m_1+ n_2m_2) &
\nonumber \\
& + (n_1n_2-1)(m_1m_2-1)-1\geq 0. &
\end{eqnarray}
If this inequality is violated the state in question is
inseparable, {\it i.e.} entangled.  Note that we still have $\bar
n$ equal for the two modes of the environment.

As a special case, we prove that when one mode of the entangled state
decoheres into a phase-sensitive environment and the other mode into
a phase-insensitive environment, the lifetime of entanglement does not
get lengthened compared to the case when both the modes decohere to
phase-insensitive environments.  We know that $\delta$ in 
Eq.(\ref{OurSimonsCriterion}) is a monotonously increasing function between
$0\leq r^2\leq 1$ as entanglement is not increased by interacting with
environments.  Thus our task is to show
\begin{equation}
\label{simcritthermalvssq}
\delta(s_{e1}=s_{e2}=0) \leq \delta(s_{e1}\neq 0, s_{e2}=0).
\end{equation}
By substituting Eq.(\ref{19}) into Eq.(\ref{OurSimonsCriterion}), 
\begin{eqnarray}
& \delta(s_{e1}\neq 0, s_{e2}=0) - \delta(s_{e1}=s_{e2}=0) & \nonumber \\
& =4r^2t^2\tilde{n}\sinh^2s_{e1}[(\mu m_1-\lambda^2 t^2)m_2-\mu] &
\end{eqnarray}
As $4r^2t^2\tilde{n}\sinh^2s_{e1}\geq 0$,  $E\equiv(\mu m_1-\lambda^2 t^2)m_2-\mu$
will determine the sign.  Using Eq.(\ref{19}), we find that
\begin{equation}
\label{20}
E=(\mu\tilde{n}-1)(\tilde{n}-\mu)r^4+(\mu^2\tilde{n}-2\mu+\tilde{n})r^2.
\end{equation}
It is straightforward to show that $E$ is always positive between $0\leq r^2\leq 1$.
We have thus proved Eq.(\ref{simcritthermalvssq}) and shown that making
one mode of the environment phase-sensitive does not prolong the
lifetime of entanglement.

\begin{figure}
\centerline{\scalebox{1.0}{\includegraphics{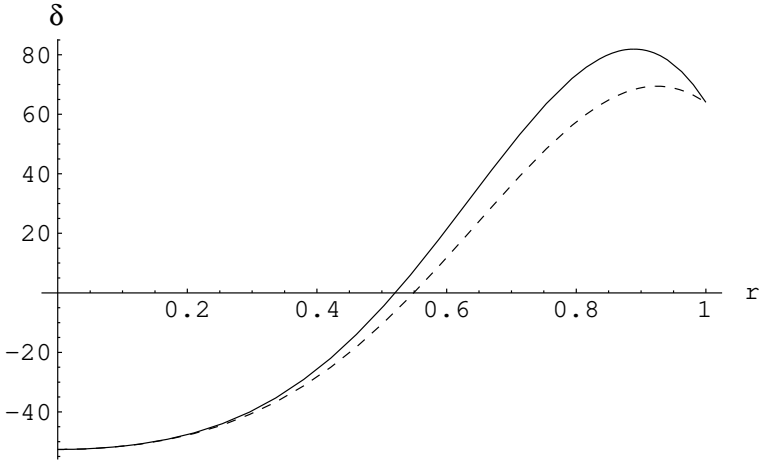}}}
\vspace{0.5cm}
\caption{Separability $\delta$ versus normalised time $r$, where
  squeezing of the channel $s_c=1$ and the mean number of photons
  $\bar{n}=1$.  The dashed line represents the evolution of the
  separability in a thermal (phase-insensitive) environment where
  $s_{e1}=0$ and $s_{e2}=0$.  The solid line represents the evolution of the
  separability in an environment one mode of which is squeezed (phase-sensitive) 
  with $s_{e1}=0.5$ and the other is phase-insensitive $s_{e2}=0$.  The coherent lifetime of the system
  in the phase-sensitive environment is less than that of the system
  in the thermal environment.}
\label{fig1}
\end{figure}

In Figure \ref{fig1} the contrast between a two-mode squeezed
channel interacting with thermal environments and a two-mode
squeezed channel interacting with an environment one mode of which
is phase-sensitive.  The squeezing of the thermal environment
results in a reduction of the coherent lifetime.

\section{Remarks}
A quantum state decoheres as it interacts with an environment.  
We have investigated the possibility of keeping entanglement longer by squeezing
the environment.    We have found that squeezing the
environment shortens the coherence lifetime of the system.  Owing to the
fact that entanglement is due to the quantum nature in both of the quadratures so that
reducing added noise into one quadrature does not lengthen the lifetime because
this action is done at the expense of stretching the added noise into the conjugate quadrature.

\acknowledgments

We would like to thank Dr. Bill Munro and Mr. H. Jeong for helpful discussions.
This work has been supported by the UK Engineering and Physical Sciences
Research Council (GR/R33304). D.W. is grateful for financial support from the
Department of Education and Libraries (DEL).

\end{document}